\documentclass{paper}

\usepackage{graphicx}  
\usepackage{dcolumn}   
\usepackage{bm}        
\usepackage{amssymb}   

\usepackage{url}

\begin{document}

\markboth{Yu. G. Gordienko}{Generalized Model of Migration-Driven Aggregate Growth ---
Asymptotic Distributions, Power Laws and Apparent Fractality}

%
%

\title{Generalized Model of Migration-Driven Aggregate Growth ---
Asymptotic Distributions, Power Laws and Apparent Fractality}

\author{Yuri G. Gordienko \\
G.V.Kurdyumov Institute for Metal Physics, \\
National Academy of Sciences of Ukraine,\\
36 Academician Vernadsky Blvd, Kiev 03680 Ukraine\\
gord@imp.kiev.ua\\
\\
\\
Received 18 August 2011\\
Revised 26 September 2011
}

\maketitle

\begin{abstract}
The rate equation for exchange-driven aggregation
of monomers between clusters of size $n$ by power-law exchange rate ($\sim{n}^\alpha$),
where detaching and attaching processes were considered separately,
is reduced to Fokker-Planck equation.
Its exact solution was found for unbiased aggregation and agreed with asymptotic conclusions of other models.
Asymptotic transitions were found from exact solution
to Weibull/normal/exponential distribution, and then to power law distribution.
Intermediate asymptotic size distributions were found to be functions
of exponent $\alpha$ and vary from normal ($\alpha=0$) through Weibull ($0<\alpha<1$) to exponential ($\alpha=1$) ones,
that gives the new system for linking these basic statistical distributions.
Simulations were performed for the unbiased aggregation model on the basis of the initial rate equation
without simplifications used for reduction to Fokker-Planck equation.
The exact solution was confirmed,
shape and scale parameters of Weibull distribution (for $0<\alpha<1$)
were determined by analysis of cumulative distribution functions
and mean cluster sizes, which are of great interest, because they can be measured in experiments
and allow to identify details of aggregation kinetics (like $\alpha$).
In practical sense, scaling analysis of \emph{evolving series} of aggregating cluster distributions
can give much more reliable estimations of their parameters than analysis of \emph{solitary} distributions.
It is assumed that some apparent power and fractal laws observed experimentally
may be manifestations of such simple migration-driven aggregation kinetics even.
\end{abstract}

\keywords{Aggregation kinetics; stochastic process; one-step process; Fokker-Planck equation; scaling;
scale-free distributions; crystal lattice defects; desktop grid.}

\section{Introduction}
Many aggregation phenomena in nature take place by exchange of
solitary monomers (particles) between their aggregates (clusters):
phase ordering,\cite{lifshitz1961kinetics,wagner1961theorie} atom deposition,\cite{zangwill1988physics}
growth and distribution of assets,\cite{ispolatov1998wealth}
city population,\cite{leyvraz2002scaling} etc.
In these aggregation processes particles can leave one
cluster and attach to another. Usually these exchange processes are described by an exchange rate kernel $K(i,\mbox{ }j)$, i.e. the rate of transfer of particles from a cluster of size $i$ (detaching event) to a cluster of size $j$ (attaching event). Generally, the rate of monomer
particle exchange between two clusters depends on their active interface
surfaces that are dependent on their sizes, morphology (line, plane, disk, sphere,
fractal, etc), probability of detaching and attaching events, etc.

\begin{figure}[t]
  \centerline{\includegraphics[width=3in]{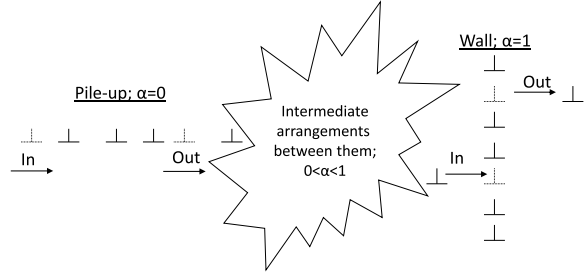}}
  \caption{Cluster arrangements with a minimum (pile-up of dislocations in this example), maximum (wall of dislocations in this example), and intermediate active surfaces. Arrows with ``In'' and ``Out'' labels show attaching and detaching events, respectively (see details in the text).}\label{Fig01}
\end{figure}

Recently, the generalized \emph{linear} model was proposed,\cite{gordienko2011change}
where detaching and attaching processes were considered \textit{separately} that in the general case
could be characterized by different rates, which differs by this aspect from
\emph{nonlinear} models in Leyvraz-Redner scaling theory of aggregate growth,\cite{leyvraz2002scaling}
Ben-Naim-Krapivsky theory for exchange driven growth,\cite{ben2003exchange}
Lin-Ke theory for migration-driven aggregation.\cite{ke2002exchange,lin2003exchange,lin2005exchange}

The different \textit{detach} product kernel $K_d (i)=k_d S_d\left( i \right)$
and \textit{attach} product kernel $K_a (i)=k_a S_a \left( i \right)$ were taken into
account, where $k_d$ and $k_a$ are the measures of activation of detaching and attaching processes.
$S_d \left( i \right)=s_d i^\alpha $ and
$S_a \left( i \right)=s_a i^\beta $
are the active surfaces of clusters,
where $\alpha $ and $\beta $ are the exponents depending on the morphology of cluster.
The probability distribution function (PDF)
or density of clusters $f(n,t)$ containing the $n$ particles at time $t$ evolves
according to the following linear rate equation:
\begin{equation}
\label{eq1}
\begin{array}{l}
 \frac{\partial f(n,t)}{\partial t} = K_d (n+1)f(n+1,t)+K_a (n-1)f(n-1,t)- \\
 -K_d (n)f(n,t)-K_a (n)f(n,t).
 \end{array}
\end{equation}
It should be noted that comparison between this linear aggregation model and
the aforementioned nonlinear models\cite{ben2003exchange,ke2002exchange,lin2003exchange,lin2005exchange}
is appropriate in the view of their final linearization.
For example, in Ben-Naim-Krapivsky theory for exchange driven growth
the moments of aggregate size distribution
were absorbed in a new time variable
to reduce the nonlinear rate equation (eq.2 in their work\cite{ben2003exchange})
and analyze the linear differential equation (eq.4 in their work\cite{ben2003exchange}) finally.

\section{Reduction of rate equation to Fokker-Planck equation}
For high values of $n$ one can get from (\ref{eq1}) the following reduced equation:\cite{gordienko2011change}
\begin{equation}
\label{eq2}
\frac{\partial f\left( {n,t} \right)}{\partial t} = \frac{\partial
\left( {D_1 \left( n \right)f\left( {n,t} \right)} \right)}{\partial
n}+\frac{\partial ^2\left( {D_2 \left( n \right)f\left( {n,t} \right)}
\right)}{\partial n^2},
\end{equation}
i.e. the Fokker-Planck equation\cite{fokker1914mittlere,planck1917uber} with
power-law drift $D_1 (n)=K_d (n)-K_a (n)= s\left( {n^\alpha k_d - n^\beta k_a }
\right)$ and diffusion $D_2 (n)={1 \over 2}(K_d (n) + K_a (n))={1 \over 2}(n^\alpha sk_d + n^\beta sk_a)$ coefficients.
Here the simplest case of unbiased aggregation is considered, i.e. when $\alpha = \beta$, and $k_d = k_a$, subsequently.

Some special cases of (\ref{eq1}) and (\ref{eq2})
are of interest for some practical applications.
\cite{lifshitz1961kinetics,wagner1961theorie,ispolatov1998wealth,leyvraz2002scaling,gordienko2011change,zangwill1988physics}
The case $\alpha=0$
corresponds to the clusters with the minimum active surface,
which is independent of the whole number of particles $n$ in it.
Such configurations similar to queues and
stacks in computer science, and to ``pile-up''
aggregation of dislocations of regular crystalline structure (Fig.\ref{Fig01}).
This case is described by the homogeneous heat equation
and leads to the ``diffusive-like kinetic universality class''.\cite{gordienko2011change}
The case $\alpha =1$
corresponds to the clusters with the maximum active surface.
For example, ``wall'' aggregation of dislocations of
regular crystalline structure can be depicted by this scenario (Fig.\ref{Fig01}).
This case leads to the ``ballistic (exponential) kinetic universality class''.\cite{gordienko2011change}
It should be noted that these two partial cases were considered also
in Lin-Ke theory for migration-driven aggregation,
namely for $\alpha=0$ in their work\cite{ke2002exchange}
and for $\alpha=1$ in their other work\cite{lin2003exchange},
but starting from other assumptions and formulations (see below additional comments as to comparison with their results).

\section{Exact asymptotic solution of the reduced Fokker-Planck equation}
Here the more general case $0<\alpha<1$ is considered,
which is related to clusters, which bulk particles shielded by
active surfaces (for example, perimeter of disk, fractal, etc).
In solid state physics, such configurations can take place in
various arrangements of defects of crystalline structure
(from compact voids to spare fractals).\cite{zaiser2004self,hahner1998fractal,gordienko1994synergetic,gordienko1996metastable}
For $k_d =k_a =k$ and $D=sk$ one can obtain:

\begin{equation}
\label{eq3}
\frac{\partial f\left( {n,t} \right)}{\partial t}=D\frac{\partial ^2\left(
{n^\alpha f\left( {n,t} \right)} \right)}{\partial n^2}.
\end{equation}

\begin{figure}[hb!]
    \centerline{\includegraphics[width=2.8in,angle=270]{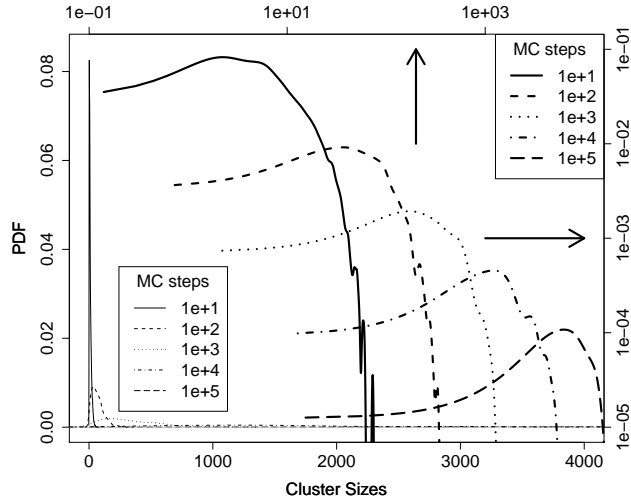}}
\caption{\label{Fig02} PDFs for $\alpha=1/2$ in linear (left group of thin curves) and log-log (right group of thick curves) coordinates.}
\end{figure}

Below the time evolution of the initial singular
distribution of $N_0 $ clusters of the same size $n_0 $ that exchange by
particles is considered. It is the first boundary value problem for
a domain $0<n<\infty$ with initial $f\left( {n,0} \right)=N_0 \delta
\left( {n-n_0 } \right)$ and boundary $f\left( {0,t} \right)=0$, $f\left(
{\infty ,t} \right)=0$ conditions. After substitution $g\left( {n,t}
\right)=n^\alpha f\left( {n,t} \right)$ one can get the simplified equation,
which is similar to the equation of diffusion in a turbulent medium.\cite{sutton1943equation,liu2006turbulent}

Using the method of separation of variables (\ref{eq3}) was reduced to the
set of ordinary and Bessel differential equations,\cite{jahnke1945tables}
and then by integral transformation on
the basis of Weber's second exponential integral,\cite{watson1952treatise}
Green's function was constructed and the exact solution was found:

\begin{equation}
\label{eq4}
f(n,t)=\frac{N_0 n^{1 \over 2}n_0^{1 \over 2}}
{n^{2-p}pDt}\exp \left[ {-\frac{n^p+n_0^p}{p^2Dt}}
\right]I_{1 \over p}\left({\frac{2n^{p \over 2}n_0
^{p \over 2}}{p^2Dt}}\right),
\end{equation}

where $p=2-\alpha$ and $I_{1/p}$ is a modified Bessel function.
For $t\gg{t_W}={2n^{p/2}n_0^{p/2}}/{p^2D}$
a modified Bessel function becomes close to
power function faster than exponential one goes to unity
and the solution (\ref{eq3})
$f_{W}(n,t)=f(n\gg{n_0},t\gg{t_W})$ will be close to:

\begin{equation}
\label{eq5}
f_{W} (n,t)=\frac{N_0 n_0 n^{p-1}}
{\Gamma({1 \over p}) p^{2 \over p} (Dt)^{{p+1} \over p} }
\exp \left[ {-\frac{n^p}{p^2Dt}} \right]\;,
\end{equation}

where $\Gamma(1/p)$ is the gamma function.

It should be noted that equation (\ref{eq5}) is in agreement with
results of scaling analysis on the basis of the approximate ansatz function
in Ben-Naim-Krapivsky theory for exchange driven growth,
namely with eq.(13) in their work\cite{ben2003exchange}
taking into account the rescaled time variable.

After substitution $b=\left( {p^2Dt} \right)^{1/p}$ PDF (\ref{eq5}) will be:

\begin{equation}
\label{eq6}
f_{W} (n,t)=\frac{pN_0 n_0 }{\Gamma \left( {1 \mathord{\left/
{\vphantom {1 p}} \right. \kern-\nulldelimiterspace} p} \right)b}\;W\left(
{n;b,p} \right)\;,
\end{equation}

(where $W\left( {n;b,p}
\right)=\frac{pn^{p-1}}{b^p}\exp \left[ {-\left( {n \mathord{\left/
{\vphantom {n b}} \right. \kern-\nulldelimiterspace} b} \right)^p} \right]$ is a Weibull distribution)
and its cumulative distribution function (CDF) is $cdf_{W} (n,t)=1-\exp \left[ {-\left( {n \mathord{\left/ {\vphantom {n b}}
\right. \kern-\nulldelimiterspace} b} \right)^p} \right]$, which is the exact CDF for a Weibull distribution.
These results support the previous estimations,\cite{gordienko2011change}
that in the
limit case $\alpha =0$ (minimum active surface) aggregation goes with the
classic ``diffusive'' kinetics $cdf_{W} \left(
{n,t\vert \alpha =0} \right)=1-\exp \left[ {-{n^2} \mathord{\left/
{\vphantom {{n^2} {4Dt}}} \right. \kern-\nulldelimiterspace} {4Dt}} \right]$
and in the other limit case $\alpha =1$ (maximum active surface) aggregation
goes with the known ``linear'' (or ``ballistic'') kinetics
$cdf_{W} \left( {n,t\vert \alpha =1} \right)=1-\exp
\left[ {-n \mathord{\left/ {\vphantom {n {Dt}}} \right.
\kern-\nulldelimiterspace} {Dt}} \right]$,
which is actually exponential.
For a case
$0<\alpha <1$ the intermediate ``fractional'' kinetics could be observed:

\begin{equation}
\label{eq7}
cdf_{W}(n,t)=
1-\exp \left[ -n^{2-\alpha} / (2-\alpha)^2Dt \right].
\end{equation}

Thus, asymptotic size distributions were found to be functions
of exponent $\alpha$ and vary from normal ($\alpha=0$) through Weibull ($0<\alpha<1$) to exponential ($\alpha=1$) ones.
Again, it should be noted that exact solution (\ref{eq4})
and equation (\ref{eq7}) are in agreement with
exact solutions of rate equations for two partial cases
in Lin-Ke theory for migration-driven aggregation,
namely for $\alpha=0$ in their work\cite{ke2002exchange}
and for $\alpha=1$ in their other work\cite{lin2003exchange},
and in agreement with results of scaling analysis on the basis of the approximate ansatz function
in Ben-Naim-Krapivsky theory for exchange driven growth,
namely with eq.(13)\cite{ben2003exchange}.

For later time $t{\gg}t_{pl}={n^p}/{p^2D}$ (where $t_{pl}>t_W$) the left tail of CDF (\ref{eq7})
will be close to the following power law:

\begin{equation}
\label{eq8}
cdf_{pl} ( n,t) \approx n^{2-\alpha} / (2-\alpha)^2Dt.
\end{equation}

It is assumed that some apparent power and fractal laws observed experimentally
as linear regions on double logarithmic plots of CDFs
(for example, observed on surfaces of plastically deformed single crystals\cite{gatsenko2011statistical,gordienko2008multiscale})
may be manifestations of such simple migration-driven aggregation kinetics in
ensemble of solid state defects.

\begin{figure}[hb!]
    \centerline{\includegraphics[width=2.8in,angle=270]{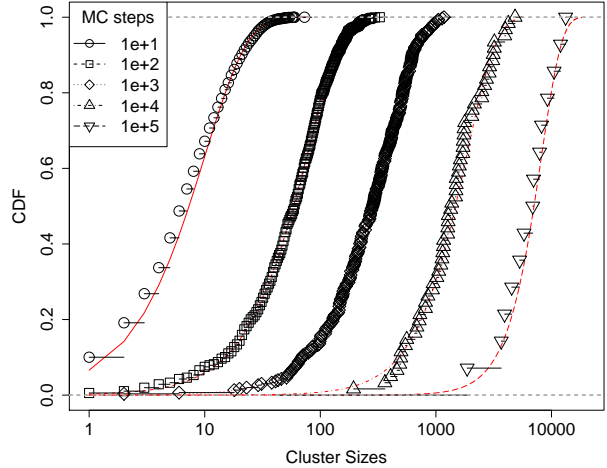}}
\caption{\label{Fig03} Simulated CDFs (symbols) for $\alpha=1/2$ fitted by theoretical Weibull distributions (lines).}
\end{figure}

\section{Simulation of unbiased aggregation with separate detaching and attaching events}
This model was tested by computer simulations
by Monte Carlo method in the distributed computing
infrastructure (DCI) ``SLinCA@Home''\cite{SLinCA}
on BOINC SZTAKI Desktop Grid (DG).\cite{Kacsuk,urbah2009edges}
Numerous initial configurations of
clusters with various initial numbers of particles $N_0$ in each cluster
were tested in simulations
with the conserved number of all aggregating particles in each run
(from $10^4$ to $10^6$).
Some PDFs (Fig.\ref{Fig02}) and CDFs (Fig.\ref{Fig03}) are shown
for initial configurations of $10^5$
clusters with 1 particle per cluster and other cases
will be reported in details separately elsewhere.
In comparison of analytical and simulated results
the number of Monte Carlo steps (MCSs)
is assumed to be equal to the number of time steps $t$.

The idea behind simulation is that
the proposed results of simulations are
not \emph{merely numerical verification} of asymptotic analytic solution,
but imitation of the one-step aggregation model itself
(in the general formulation by eq.(\ref{eq1})),
which seemed to be obligatory for comparison of the numerous rough simplifications and scaling assumptions
made in other works and exact solution in this work.

\subsection{Statistical test of simulated results}
Despite the visually good agreement (Fig.\ref{Fig03}) between simulated data (CDFs) and
the fitting curves (by Weibull distributions) the goodness of fit was checked in two tests,
where the null and alternative hypotheses were:

- H0: Data come from the stated distribution;

- HA: Data \emph{do not} come from the stated distribution,

where ``stated'' distributions were Weibull
and normal (for comparison) distributions.
The necessity for such verification is that many other sigmoidal distributions,
and not only Weibull distribution, could give an equally good visual fit.
Moreover, some distributions in the limited range of scales can be
erroneously accepted as others\cite{sornette2006critical}.
The other reason is related to a problem of reliable distribution fitting
the limited sets of experimental data.
The idea behind the statistical test of the simulated results
is to investigate the range of validity of the approximate solution in eq.(\ref{eq5}) applied for the simulated results.
It will allow to determine the actual limits of its applicability
to estimation of various experimental distributions\cite{gordienko2011change,gatsenko2011statistical}.

\begin{figure}[ht!]
    \centerline{\includegraphics[width=2.8in,angle=270]{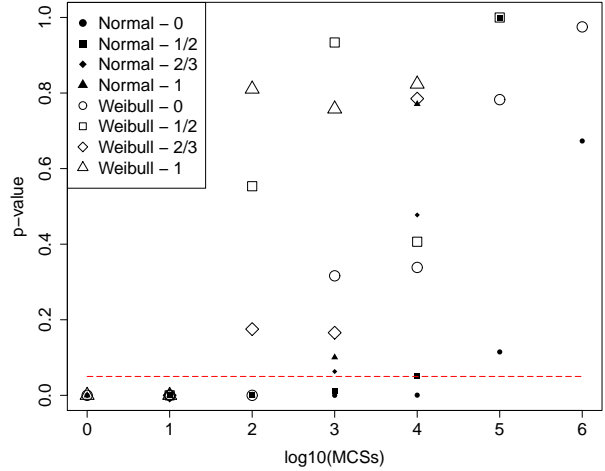}}
\caption{\label{Fig04} Results of KS-test: p-values from for simulated CDFs fitted by normal (closed symbols) and Weibull (open symbols) distributions
for $0\leq\alpha\leq1$ (in legend). The significance level of 0.05 is noted by dash line (see explanations in the text).}
\end{figure}

The Kolmogorov-Smirnov (KS) test\cite{kolmogorov1933,smirnov1948,corder2009}
was used to decide if simulated data comes from a population with Weibull
and normal distributions, which is based on a comparison between the empirical
and theoretical CDFs (Fig.\ref{Fig03}).
From Fig.\ref{Fig04} one can see that
for Weibull (but not for normal) distribution p-values are enough higher
than significance level of 0.05 (noted by dash line) usually referred in statistical literature.
It means that H0-hypothesis for Weibull distribution
(i.e. that the simulated data follow a Weibull distribution) can be accepted
for open symbols in Fig.\ref{Fig04} in the range of MCs, where p-values higher that 0.05,
namely for $t>10^2$, that agrees with the stated Weibull-like asymptotic (\ref{eq7}) for $t_{pl}>t>t_W$.
It should be noted that H0-hypothesis for normal distribution is also can be considered as valid in limited range of MCSs,
namely for $t>10^4$ MCSs, because normal and Weibull distributions cannot be distinguished for
the limited number of big clusters
($<10$ for $\alpha=1$ --- triangles; $<20$ for $\alpha=2/3$ --- diamonds; $<50$ for $\alpha=1/2$ --- squares in Fig.\ref{Fig04})
on the very late stages of aggregation,
and validity of their distribution function fitting cannot be reliably estimated by KS-test even.
That is why the more simulations should be carried out to check the range of validity of the asymptotic results
eq.(\ref{eq4}) stated in this work
and the one-step aggregation model itself in the general formulation by eq.(\ref{eq1}).

\begin{figure}[hb!]
    \centerline{\includegraphics[width=2.8in,angle=270]{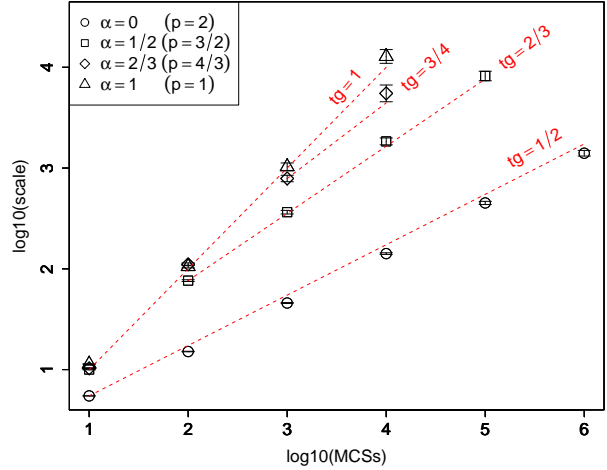}}
\caption{\label{Fig06} Scale parameters ($b$) for Weibull distributions used as fits for the simulated PDFs in Fig.\ref{Fig02}.}
\end{figure}


\subsection{Analysis of evolving distribution parameters and mean values with implications for practice}
The shape of simulated PDFs changes crucially during simulations (Fig.\ref{Fig01}):
from the initial symmetric singular (like Dirac-function) PDF
($f\left( {n,0} \right)=N_0 \delta \left( {n-n_0 } \right)$),
then to asymmetric sigmoidal PDF due to influence of boundary condition $f\left( {0,t} \right)=0$,
and then to sparse PDF with the small number of big clusters and long tails on the late stage of simulation (for $t\geq10^5$).
That is why the simulated shape ($p$) parameters roughly correspond to the theoretical values for Weibull distribution
during the whole term of simulation run.
This allow to determine the actual range of validity of the asymptotic solution in eq.(\ref{eq4}),
which actually was derived from the more complex exact solution eq.(\ref{eq3})
for narrow time range (MCs range).
In contrary, scale ($b$) (Fig.\ref{Fig06}) parameters evolve with time
in the very steady fashion and their tangents $tg=p^{-1}$ demonstrate
the excellent coincidence with analytical predictions,
which is evident from available scaling.

\begin{figure}[ht!]
    \centerline{\includegraphics[width=3.5in]{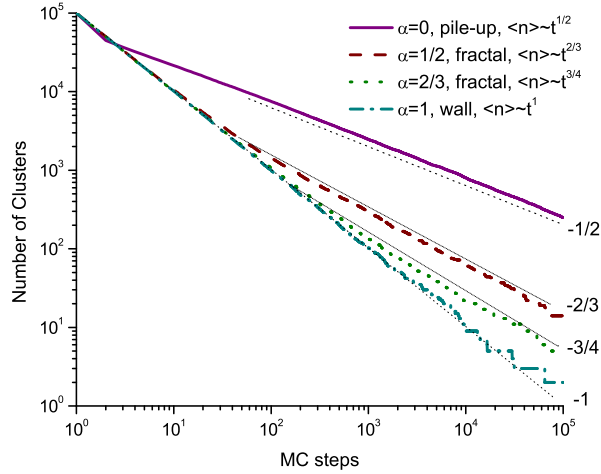}}
\caption{\label{Fig07} The simulated numbers of clusters $N$ (thick lines)
and analytical predictions $N \sim {\langle n \rangle}^{-1} \sim t^{-1/(2-\alpha)}$ (thin lines).}
\end{figure}

The average values are of great interest, because they can be
measured in experiments and not so vulnerable to fluctuations in PDFs
and mainly follow the general scaling law.
For example, from (\ref{eq5}) the mean cluster size should grow like $\langle {n} \rangle = b/p~\Gamma(1/p)\sim t^{1/(2-\alpha)}$, and it is confirmed by simulations for various values of $0\leq\alpha\leq1$ (Fig.\ref{Fig07}).
From the exact solution (\ref{eq4}) the mean cluster size is equal to:
\begin{equation}
    \label{eq9}
    \langle {n} \rangle = n_0 \frac{\Gamma(1/p)}{ \Gamma(1/p) - \Gamma \left( 1/p, \frac{{n_0}^p}{p^2Dt} \right) },
\end{equation}
which is also goes to  $t^{1/(2-\alpha)}$ for $t\gg{n_0}^p/(p^2D)$.
It is in agreement with conclusions on average cluster size for two partial cases
in Lin-Ke theory for migration-driven aggregation,
namely for $\alpha=0$ (diffusive growth\cite{ke2002exchange})
and $\alpha=1$ (ballistic growth\cite{lin2003exchange}),
and estimations of the typical scale growth for diffusive and ballistic regimes
in Ben-Naim-Krapivsky theory for exchange driven growth.\cite{ben2003exchange}
It is also in agreement with conclusions of nonlinear Leyvraz-Redner scaling theory\cite{leyvraz2002scaling}
(with the conserved number of monomers)
about mean aggregate growth with time. The other moments can be calculated similarly
and these results will be reported in detail separately elsewhere.
In practical sense, it means that scaling analysis of \emph{evolving series} of aggregating cluster distributions (especially, their CDFs, mean values, and other moments) can give much more reliable estimations of their parameters than analysis of \emph{solitary} PDFs or CDFs.

\begin{figure}[hb!]
    \centerline{\includegraphics[width=2.2in,angle=270]{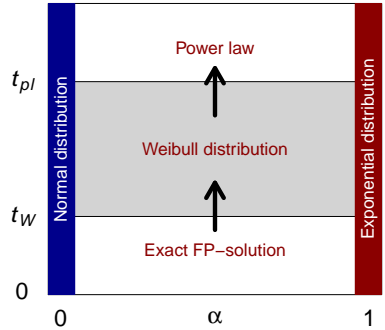}}
\caption{\label{Fig08} System of distributions and asymptotic transitions.}
\end{figure}

\section{Fokker-Planck equation of migration-driven aggregation and system of distributions}
This model allows to propose the new system of distributions in addition to Pearson
and, especially, Burr systems.\cite{burr1942cumulative}
The matter is by change of a parameter ($\alpha=0$) in FP-equation
with power-law diffusion coefficient (\ref{eq2})
one can generate family of distributions by asymptotic FP-solutions for $t_{pl}>t>t_W$:
from normal ($\alpha=0$) through Weibull ($0<\alpha<1$)
to exponential ($\alpha=1$) distribution (along horizontal axis in Fig.\ref{Fig08}).
The further extension of this system with inclusion of other distributions (for example, gamma instead of Weibull)
could be made after taking account other FP-solutions of Eq.(\ref{eq2}) with a non-zero drift coefficient.
In addition, gradual transitions with time were found for size distributions in ensembles of aggregating monomers:
from the exact FP-solution (including power, exponential, and Bessel functions)
to Weibull distribution, and then to power law distribution
(along vertical axis in Fig.\ref{Fig08}).
It is assumed, that in experiments such size distributions of aggregating monomers
at late stage of their evolution could be estimated as ones obeying power or fractal laws
(like apparent power laws and fractality in crystal defect structures\cite{zaiser2004self,hahner1998fractal,gatsenko2011statistical,gordienko2008multiscale,zasimchuk2003equidimensional}).

\section{Conclusion}
Finally, the model of exchange-diven aggregation kinetics of monomers in clusters is proposed on the basis of \emph{separate}
detaching and attaching events,
those in the general case could be characterized by \emph{different}
power-law exchange (detach and attach) rates ($\sim{n}^\alpha$) as functions of cluster size $n$.
In an asymptotic regime for high values of $n$
it allows us to construct the linear rate equation for exchange-driven aggregation kinetics,
which can be reduced to a Fokker-Planck equation with power-law drift and diffusion coefficients.
Exact asymptotic solution of the Fokker-Planck equation for unbiased aggregation (without drift term) was obtained,
that agrees with asymptotic conclusions of \emph{nonlinear} models
(those are actually become \emph{linear} under condition of the conserved number of monomers)
in Leyvraz-Redner scaling theory of aggregate growth,\cite{leyvraz2002scaling}
Ben-Naim-Krapivsky theory for exchange driven growth,\cite{ben2003exchange}
Lin-Ke theory for migration-driven aggregation.\cite{ke2002exchange,lin2003exchange,lin2005exchange}
In relation to practical applicability of these exact asymptotic results,
the kinetic Monte Carlo simulations were carried for the unbiased aggregation model on the basis the initial rate equation
without linearization and simplifications used for reduction to Fokker-Planck equation.
The simulation results confirm the exact solution in the range of the used asymptotic assumptions,
that was validated by statistical Kolmogorov-Smirnov test.
It is shown that fitting and scaling analysis of evolving CDFs and mean values of distributions
allow to determine some intrinsic features of empirical distributions
(for example, diffusive and ballistic kinetic universality classes,
and distinguish different morphologies of aggregating clusters in this research).

Asymptotic transitions in cluster size distributions (from normal to Weibull and to exponential)
are found to be dependent on the value of exponent $\alpha$ in power-law exchange (detach and attach) rates ($n^\alpha$).
In the more general statistical context this allows us to propose
the new way for systematization of distributions (in addition to Pearson and Burr systems)
that can linked by Fokker-Planck equation with power-law diffusion coefficient.
It is assumed, that some experimental power/fractal laws in size distributions of aggregating monomers
at late stage of their evolution can be created by such simple kinetics even.

\section*{Acknowledgments}
Author is grateful to anonymous reviewers of the preprint of this
paper\cite{gordienko2011generalized} for their discussion with useful comments and critics. The work was partially funded by the FP7 DEGISCO
(Desktop Grids for International Scientific Collaboration) (http://degisco.eu) project supported by the FP7
Capacities Programme, agreement number RI-261561.

\end{document}